\documentclass[12pt,a4paper]{article}

\usepackage[utf8]{inputenc}
\usepackage[T1]{fontenc}
\usepackage[margin=2.5cm]{geometry}
\usepackage[numbered]{bookmark}
\usepackage{mathtools}
\usepackage{amssymb}
\usepackage{graphicx}
\usepackage[font=small,labelfont=bf]{caption}
\usepackage{authblk}
\usepackage{cite}
\usepackage{dsfont}
\usepackage{xcolor}
\usepackage[titles]{tocloft}

\numberwithin{equation}{section}
\allowdisplaybreaks

\newcommand{\eq}[1]{\begin{equation} #1 \end{equation}}
\newcommand{\al}[1]{\begin{align} #1 \end{align}}

\newcommand{\pa}{\partial}

\newcommand{\ket}[1]{|#1\rangle}
\newcommand{\bra}[1]{\langle#1|}
\newcommand{\braket}[1]{\langle#1 \rangle}

\begin{document}

\title{\vspace{2cm}\textbf{Notes on entanglement entropy for excites holographic states in 2d}}
\author[*,$\dagger$]{Radoslav C. Rashkov}
\affil[*]{\textit{Department of Physics, Sofia University,}\authorcr\textit{5 J. Bourchier Blvd., 1164 Sofia, Bulgaria}\vspace{5pt} \authorcr{ and} \vspace{5pt}} 
\affil[$\dagger$]{\textit{Institute for Theoretical Physics, Vienna University of Technology,}\authorcr\textit{Wiedner Hauptstr. 8--10, 1040 Vienna, Austria}
	\authorcr\vspace{10pt}\texttt{rash@hep.itp.tuwien.ac.at}
	\vspace{1cm}}
\date{}

\maketitle

\begin{abstract}
In this work  we revisit the problem of contributions of excited holographic states to the entanglement entropy  in two-dimensional conformal field theories. Using the results of replica trick method we find three expressions for these contributions. First, we express  the contribution of the excited states in terms of Aharonov invariants. It is shown that beside the Schwarzian, the one-point functions of descendants of energy-momentum also contribute. Given Schwarz-Christoffel map, the contributions to any order can be easily computed. The second expression relates the entanglement entropy of excited states to Faber polynomials and Grunsky coefficients. Based on the relation of Grunsky coefficiens to tau-funcion of dispersionless Toda hierarchy, we find the third expression for contributions of excited holographic states to the entanglement entropy.
\end{abstract}

\vspace{1.5cm}

\noindent\rule{\linewidth}{0.75pt}
\tableofcontents
\noindent\rule{\linewidth}{0.75pt}

\section{Introduction}

Recent progress in holographic correspondence uncovered remarkable relations between entanglement entropy (EE) and a specific extremal surface area. From the information space point of view however, this property is not expected to be universal. For instance, as pointed in \cite{Beach:2016ocq}, the space of entanglement entropies  is much larger than the space of asymptotically AdS metrics. One can consider geometries dual to holographic states instead. This approach assumes that the bulk content represents, in possibly quite non-trivial way, the evolution of particular boundary data. The latter basically consists of correlation functions of certain local operators. Thus, studying some non-local properties like entanglement entropy in this context is interesting and challenging task.

There is a number of papers on holographic entanglement entropy (and not only!) marking a tremendous success in our understanding of these issues. Here we will mention only several of them, used directly for our findings.

In this short note we find that entanglement entropy of excited states, calculated via replica-trick method, has an expansion in terms of so-called Aharonov invariants. The first one is pre-Schwarzian and the second is the Schwarzian itself. This confirms the result of \cite{Beach:2016ocq} that one-point function of the energy-momentum is related to the entanglement entropy. However, we find that EE contains also derivarives of the one-point function. Next, we find that once we have found the so-called Grunsky coefficients (the cofficients of Faber polynomials), we can calculate the expansion of entanglement entropy to given order (in fact to all orders). Having the fact that Grunsky coefficients are second derivatives of dispersionless Toda (dToda) tau-function, we conjecture that entanglement entropy is intimately related to Toda hierarchy. 

In the next section we briefly review the derivation of Renyi entropy for excited states using the replica trick method. Then we make use of the expansion in terms of Aharonov invariants and comment on the role of the one-point function of the energy-momentum tensor.  Next section considers Faber polynomials, Grunsky coefficients and their relation to Schwarzian. Then we briefly demonstrate the relation of Grunsky coefficients to tau-function of dispersionless Toda hierarchy. We conclude with brief comments on our findings.


\section{Entanglement entropy for excited states and Aharonov invariants}

The M\"obius transformations (or its universal cover $SL(2,\mathbb{C}$) leave the vacuum invariant. The vacuum entanglement entropy has been considered in many papers, see for instance [Cardy-Calabrese]. An efficient way to account for contributions of excited states is to perform conformal transformations. Indeed, in general the vacuum state transforms into excited states, while the M\"obius subgroup preserves the vacuum. 

The action of the full conformal group includes the full Virasoro algebra. The identity operator and its descendant - the stress tensor (quasi-primary), participate by arbitrary products and derivatives of the stress tensor
\eq{
	\mathrm{Id}\sim 1, T, \pa^k T, T^2, T\pa^l T, \cdots.
}
According to the holographic conjecture these are the states that capture the gravitational sector of the dual theory. In the same way one can consider primary operators and their descendants to obtain other excited states. 

To briefly sketch the procedure suggested in \cite{Beach:2016ocq}, let us assume that a conformal transformation $z\:\mapsto\:w=f(z)$ is realized by an unitary operator $U_f$ ($U_fU_f^\dagger=1$). Consider the primaries $\Phi_\pm$ with the transformation properties 
\eq{
	\braket{\Phi_+(z_1)\Phi_-(z_2)}=\left(\frac{\pa f(z_1)}{\pa z_1}\right)^{h_n} \left(\frac{\pa f(z_1)}{\pa z_1}\right)^{h_n} \braket{\Phi_+(f(z_1))\Phi_-(f(z_2))}.
	\label{transf-twist}
}

The detailed calculation of the R\'{e}nyi entropy was described in \cite{Calabrese:2004eu} and here we quote their result.
There the authors have shown that, using the replica trick method, the R\'{e}nyi entropy for the vacuum is given by
\eq{
	\exp\left((1-n)S^{(n)}\right)=\braket{\Phi_+(z_1)\Phi_-(z_2)}=\frac{1}{(z_1-z_2)^{2h_n}},
}
where twist operators $\Phi_\pm(z)$ have dimensions $(h_n,\bar{h}_n)=c/24(n-1/n,n-1/n)$.
On other hand, the entanglement entropy is derived by taking the limit $n\to 1$ of the R\'{e}nyi entropy $S^{(n)}$
\eq{
	S_{vac}=\lim\limits_{n\to 1}S^{(n)}=\lim\limits_{n\to 1}\log(z_1-z_2)^{-2h_n}=\frac{c}{12}\log\frac{(z_1-z_2)}{\delta^2}.
	\label{vac-renyi-1}
}

Consider now the case of excited states $\ket{f}=U_f\ket{0}$ (the vacuum is invariant under global  $SL(2,\mathbb{Z})$ transformations). The calculation of the R\'{e}nyi entropy goes analogously
\al{
	\exp\left((1-n)S^{(n)}_{ex}\right)&=\bra{f}\Phi_+(z_1)\Phi_-(z_2)\ket{f} =
	\braket{0|U_f^\dagger \Phi_+(z_1)U_f\:U_f^\dagger\Phi_-(z_2)U_f|0}	 \label{1-point-1}\\
	&=\left(\frac{df}{dz}\right)_{z_1}^{-h_n}\left(\frac{df}{dz}\right)_{z_2}^{-h_n} \left(\frac{d\bar{f}}{d\bar{z}}\right)_{\bar{z}_1}^{-\bar{h}_n}  \left(\frac{d\bar{f}}{d\bar{z}}\right)_{\bar{z}_2}^{-\bar{h}_n}
	\bra{0}\Phi_+(f(z_1))\Phi_-(f(z_2))\ket{0},\label{1-point-2}
}
and, taking into account \eqref{vac-renyi-1} we find 
\eq{
	S_{ex}=\lim\limits_{n\to 1}S^{(n)}_{ex}=\frac{c}{12}\log\left|\frac{f'(z_1)f'(z_2)}
	{(f(z_1)-f(z_2))^{2}} \right|.
	\label{ex-renyi-1}
}
Thus, the difference between vacuum entanglement and that of excited states is
\al{
	S_{vac}-S_{ex}&=\frac{c}{12}\log\left|\frac{f'(z_1)f'(z_2)\bar{f}'(\bar{z_1})\bar{f}'(\bar{z_2}) (z_1-z_2)^{2}}{(f(z_1)-f(z_2)^2(\bar{f}(\bar{z}_1)-\bar{f}(\bar{z}_2))^2} \right|
	\nonumber \\
	& =\frac{c}{12}\log\left|\frac{f'(z_1)f'(z_2)\bar{f}'(\bar{z_1})\bar{f}'(\bar{z_2}) (f^{-1}(w_1)-f^{-1}(w_2))^{2}}{(f(z_1)-f(z_2)^2(\bar{f}(\bar{z}_1)-\bar{f}(\bar{z}_2)^2} \right|	.
	\label{ex-contrib-1}
}


Direct calculations show that ($f'(z)\neq 0$) the expansion about $z$ is
\begin{equation}\label{kernel-1}
\frac{f'(z)f'(w)}{\left(f(z)-f(w)\right)^2}=\frac{1}{(z-w)^2}+\frac{1}{6}S(f)(z)+\frac{1}{12}S'(f)(z)(z-w)+\cdots
\end{equation}
where $S(f)$ denotes the Schwarzian derivative.

More generally,  let $f$ be a nonconstant meromorphic function on a domain $D$ in the complex plane. For $z\in D$ with $f(z)\neq\infty$, $f'(z)\neq 0$, we consider the quantity
\eq{
	G(\zeta,z)=\frac{f'(z)}{f(\zeta)-f(z)},
	\label{aharon-1}
}
which can be expanded in power series
\eq{
		G(z+w,z)=\frac{1}{w}-\sum\limits_{n=1}^\infty\psi_n[f](z)w^{n-1}.
		\label{aharon-2}
	}
The quantities $\psi_n[f](z)$ are called Aharonov invariants \cite{aharonov}.
	
One can take derivative of \eqref{aharon-1} with respect to $\zeta$ and find
	\eq{
		\frac{\pa G(\zeta,z)}{\pa\zeta}=-\frac{f'(z)f'(\zeta)}{(f(\zeta)-f(z))^2}=-\frac{1}{(\zeta-z)^2}- \sum\limits_{n=2}^\infty(n-1)\psi_n[f](z)(\zeta-z)^{n-2}.
		\label{aharon-3}
	}
The quantity $\pa G(\zeta,z)/\pa\zeta$ is invariant under M\"{o}bius transformations $M$, $M\circ f(z)$
and thus, 
		\eq{
			\psi_n[M\circ f]=\psi_n[f], \qquad n\geq 2.
		}
The quantities $(G(\zeta,z),\pa_\zeta G,\psi_n[f])$ are defined even at $f(z)=\infty$ as long as $f$ is locally univalent at $z$.

The difference between vacuum entanglement and the entanglement entropy of excited states \eqref{ex-contrib-1} takes the form
\eq{
	S_{vac}-S_{ex}=\frac{c}{12}\log\left(1+	\sum\limits_{n=2}^\infty(n-1)\psi_n[f](z)(\zeta-z)^{n}.
	\right).
	\label{schwarzian-aharonov-fin}
}
		
The first two invariants are $\psi_1[f]=\frac{1}{2}\frac{f''(z)}{f'}$ and $\psi_2[f]=\frac{1}{3!}S(f)$.
Aharonov \cite{aharonov} (see also \cite{meira,harmelin}) proved the recursion formula:
		\eq{
				(n+1)\psi_n[f]=\psi_{n-1}[f]'+\sum\limits_{k=2}^{n-2}\psi_k[f]\psi_{n-k}[f], \quad n\geq 3.
				\label{aharon-5}
			}
For instance, next few invariants are
			\eq{	
				\psi_3[f]=\frac{S(f)'}{4!}; \quad \psi_4=\frac{S''(f)}{5!}+\frac{S^2(f)}{5(3!)^2}; \quad \psi_5= \frac{S'''(f)}{6!}+\frac{3S(f)S'(f)}{6!}.
				\label{aharon-psi345}
			}
In \cite{meira} it was proven that all M\"obius invariants are derivable from the Schwarzian derivarive $S(f)$.

The main result in this section is that the contributions of the excited states to the entanglement entropy are given by Aharonov invariants $\psi_n[f]$. The latter are polynomials of the Schwarzian and its derivatives.	 


\section{Entanglement entropy, Faber polynomials and Grunsky coefficients}

In this section we introduce Faber polynomial and the corresponding Grunsky coefficients \cite{duren,Teo:2003aw}. To this end we introduce two classes of univalent functions
\al{
	& \tilde{S}=\left\{f(z)=a_1z+a_2z^2+a_3z^3+\cdots=\sum\limits_{n=1}^\infty a_nz^n,\: a_1\neq 0 \right\}
	\label{faber-16a} \\
	& \Sigma=\left\{g(z)=z+b_0+\frac{b_1}{z}+\cdots=bz+\sum\limits_{n=0}^\infty b_nz^{-n} \right\}
	\label{faber-16b}
}
Let $f\in\tilde{S}$ is univalent in a neighbourhood $V$ of $0$ and $g\in\Sigma$ is univalent in a neighbourhood $U$ of $\infty$. It is said that the pair $(f,g)$ are disjoint relative to $(V,U)$ if the sets $f(V)$ and $g(U)$ are disjoint. 
Thus, the functions
\eq{ \log\frac{g(z)-g(\zeta)}{z-\zeta}, \qquad \log\frac{g(z)-f(\zeta)}{z-\zeta}, \qquad \log\frac{f(z)-f(\zeta)}{z-\zeta},
\label{univalent-0}
}
are analytic in $U\times U$, $U\times V$ and $V\times V$ respectively.
Hence, one can expand $g'(z)/(g(z)-w))$ in series with respect to $z$ at $\infty$
 \eq{
 		\frac{g'(z)}{g(z)-w}=\sum\limits_{n=0}\Phi_n(w) z^{-n-1}, \qquad \Phi_0(w)\equiv 1.
 		\label{faber-2}
 	}

The polynomials $\Phi_n(w)$
\eq{
	\Phi_n(w)=\sum\limits_{m=0}^nb_{n,m}w^m,
}	
 are called Faber polynomials. They satisfy the recursion relation
\eq{
	\Phi_{n+1}(w)=(w-b_0)\Phi_n(w)-\sum\limits_{k=1}^{n-1}b_{n-k}\Phi_k(w)+(n+1)b_n.
	\label{faber-5}
}
The coefficients $b_{nm}$ are called Grunsky coefficients. They are symmetric in its indices and polynomials in $b_k$’s.
The Grunsky coefficients of the expansions of \eqref{univalent-0} about $(\infty,\infty)$, $(\infty,0)$ and $(0,0)$ are
\begin{align}
& \log\frac{g(z)-g(\zeta)}{z-\zeta}= -\sum\limits_{m,n=1}^\infty b_{mn}z^{-m}\zeta^{-n}, \label{faber-17a} \\
& \log\frac{g(z)-f(\zeta)}{z-\zeta}=-\sum\limits_{m=1,n=0}^\infty b_{m,-n}z^{-m}\zeta^{n}, \label{faber-17b} \\
& \log\frac{f(z)-f(\zeta)}{z-\zeta}=-\sum\limits_{m=0,n=0}^\infty b_{-m,-n}z^{m}\zeta^{n}. \label{faber-17c}
\end{align}
As a consequence, taking $\zeta=0$ in \eqref{faber-17b} and \eqref{faber-17c} one finds:
\eq{
	\log\frac{g(z)}{z}=-\sum\limits_{m=1}^\infty b_{m,0}z^{-m}, \qquad \log\frac{f(z)}{z}-\sum\limits_{m=0}^\infty b_{-m,0}z^{m},
	\label{faber-18}
}
with $b_{00}=-\log a_1$.

Let us go back to \eqref{aharon-3} which enters the entanglement entropy for excited states \eqref{ex-renyi-1}. Based on the above expressions, one can write the non-singular part as an expansion with Grunsky coefficients. Indeed,
having defined Grunsky coefficients, one can write \eqref{kernel-1} as \cite{Wiegmann:1999fr}
\eq{\label{kernel-2}
\frac{f'(z)f'(w)}{(f(z)-f(w))^{2}}\,-\,\frac{1}{(z-w)^{2}}=\frac{\partial ^{2}}{\partial z\partial w}\log \frac{f(z)-f(w)}{z-w}=-\sum _{m,n\geq 1}mn\,b_{mn}z^{-m-1}w^{-n-1}.
}
Therefore, the difference between the vacuum entanglement entropy and that of excited states
\eqref{ex-contrib-1} can be written as
\eq{
	S_{vac}-S_{ex}=\frac{c}{12}\log\left(1+(z-w)^2
	\sum _{m,n\geq 1}mn\,b_{mn}z^{-m-1}w^{-n-1}
	\right).
	\label{schwarzian-grunsky-fin}
}

The direct calculation \eqref{kernel-1} allows to find the Schwarzian in terms of Grunsky coefficients 
\eq{
\label{kernel-4}
\frac{1}{6}S(f)(z)=-\frac{1}{z^2}\sum _{m,n\geq 1}mn\,b_{mn}z^{-n-m}.	
	}
Moreover, all the Aharonov invariants can be expanded in power series with Grunsky coefficients! For instance
\eq{
\psi_3[f]=\frac{1}{4z^2}\sum mn(n+m)b_{mn}z^{-n-m-1}, \quad \psi_4[f]=
\frac{1}{5}\left[\psi_3'[f] + \left(\sum mnb_{mn} z^{-m-n}\right)^2 \right].
\label{grunsky-aharonov-1}
	}

The main result in this section is the relation of Grunsky coefficients to Aharonov invariants (and thus, to entanglement entropy of excited states) through $\psi_2[f]=\frac{1}{3!}S(f)$, the recursion formula
\eqref{aharon-5} and \eqref{kernel-4}.


\section{Relation to dispersionless Toda tau-function}

Sato approach to integrable hierarchies uses the formalism of psedodifferential operators, $\mathcal{L}=\pa+ \sum_{i=1}^{\infty}u_i\pa^{-i+1}$. Then
the dispersionless Toda hierarchy is defined as the system of differential equations
\al{
	& \frac{\pa\mathcal{L}}{\pa t_n}=\{\mathcal{B}_n,\mathcal{L}\}, \quad
	\frac{\pa\mathcal{L}}{\pa t_{-n}}=\{\mathcal{\bar{B}}_n,\mathcal{L}\}, \label{dtoda-zakh-1}\\
	& \frac{\pa\mathcal{\bar{L}}}{\pa t_n}=\{\mathcal{B}_n,\mathcal{\bar{L}}\}, \quad
	\frac{\pa\mathcal{\bar{L}}}{\pa t_{-n}}=\{\mathcal{\bar{B}}_n,\mathcal{\bar{L}}\}, \label{dtoda-zakh-2}
}
where $\mathcal{L}$ and $\mathcal{\bar{L}}$ are generating functions of unknowns  $u_i =
u_i(t;s)$, $\bar{u}_i=\bar{u}_i(t;s)$. Instead of using pseudodifferential operators, one can work in terms of their symbols
\eq{		\mathcal{L}=p+u_1+u_2p^{-1}+u_3p^{-2}+\cdots, \qquad  
	\mathcal{\bar{L}}=\bar{u}_0p^{-1}+\bar{u}_1+\bar{u}_2p+\bar{u}_3p^2+\cdots 
	\label{dtoda-4}
}
The operators $B_n$, $\bar{B}_n$ are defined by
\eq{ 
	\mathcal{B}_n = \left(\mathcal{L}^n \right)_{\geq0}, \qquad
	\mathcal{\bar{B}}_n = \left(\mathcal{\bar{L}}^{-n}\right)_{\le 0}.
}
where the truncation operations $()_{\geq 0}$ and $()_{\le 0}$ denote the
polynomial part and the negative degree part in $p$. 
The Poisson brackets for symbols are
\eq{
	\{f,g\}=p\frac{\pa f}{\pa p}\frac{\pa g}{\pa t_0}-p\frac{\pa g}{\pa p}\frac{\pa f}{\pa t_0}.
}

The free energy of dispersionless Toda system is related to tau function as $\mathcal{F}=\log\tau_{dToda}$ and from \eqref{dtoda-4}  one gets
\eq{ 
	p^m = \mathcal{L}^m-u_{m,0} \mathcal{L}^{m-1}-(u_{m,1}-u_{m-1,1}u_{m,0})\mathcal{L}^{m-2}-\cdots.
}
Following \cite{Teo:2003aw,Wiegmann:1999fr}, one can prove the relations
\al{
	\log p& =\log\mathcal{L}-\sum\limits_{m=1}^\infty\frac{1}{m} \frac{\pa^2\mathcal{F}}{\pa t_0\pa t_m}\mathcal{L}^{-m} \label{toda-tau-2} \\
	\left(\mathcal{L}^n \right)_{\geq 0}& =\mathcal{L}^n- \sum\limits_{m=1}^\infty\frac{1}{m} \frac{\pa^2\mathcal{F}}{\pa t_n\pa t_m}\mathcal{L}^{-m} =
	\frac{\pa^2\mathcal{F}}{\pa t_0\pa t_n}-\sum\limits_{m=1}^\infty\frac{1}{m} \frac{\pa^2\mathcal{F}}{\pa t_{-m}\pa t_n}\mathcal{\bar{L}}^m \label{toda-tau-3} \\
	\log p& =\log\mathcal{\bar{L}}+\frac{\pa^2\mathcal{F}}{\pa t_0^2}-\sum\limits_{m=1}^\infty\frac{1}{m} \frac{\pa^2\mathcal{F}}{\pa t_{-m}\pa t_0}\mathcal{\bar{L}}^{m} \label{toda-tau-4} \\
	\left(\mathcal{\bar{L}}^{-n} \right)_{\le 0}& = - \sum\limits_{m=1}^\infty\frac{1}{m} \frac{\pa^2\mathcal{F}}{\pa t_{-n}\pa t_{m}}\mathcal{L}^{m}
	=\mathcal{\bar{L}}^{-n}+ \frac{\pa^2\mathcal{F}}{\pa t_0\pa t_{-n}}- \sum\limits_{m=1}^\infty\frac{1}{m} \frac{\pa^2\mathcal{F}}{\pa t_{-n}\pa t_{-m}}\mathcal{\bar{L}}^{-m} \label{toda-tau-5}.
}
Now we can identify $p$ with $w=g(z)$ belonging to the class $\Sigma$ and $\mathcal{L}$ with $z$. On the other hand $\mathcal{F}(p)$ is identified with $z=G(w)$, i.e. the inverse of $g(z)$. Thus, $\mathcal{B}_n$ become Faber polynomials! Reversing the considerations and following for instance \cite{Carroll:1995mu}, one can prove that the so defined tau function satisfies the Hirota dipsersionless equation.

Comparing \eqref{toda-tau-2} with \eqref{faber-18} we find that the Grunsky coefficients $b_{nm}$ of the pair $(g=w(\mathcal{L}),\: f=w(\mathcal{\bar{L}}))$  are related to the tau function, or free energy as follows:
\al{
	& b_{00}=-\frac{\pa^2\mathcal{F}}{\pa t_0^2}, \quad b_{n,0}= \frac{1}{n}\frac{\pa^2\mathcal{F}}{\pa t_0\pa t_n}, \quad b_{-n,0}= \frac{1}{n}\frac{\pa^2\mathcal{F}}{\pa t_0\pa t_{-n}},\quad n\geq 1 \nonumber \\
	& b_{m,n}= -\frac{1}{mn}\frac{\pa^2\mathcal{F}}{\pa t_m\pa t_n}\qquad b_{-m,-n}= -\frac{1}{mn}\frac{\pa^2\mathcal{F}}{\pa t_{-m}\pa t_{-n}}, \quad n,m\geq 1\label{tau-grunsky} \\
	& b_{-m,n}= b_{n,-m}=-\frac{1}{mn}\frac{\pa^2\mathcal{F}}{\pa t_{-m}\pa t_n},\quad n,m\geq 1. \nonumber
}

Therefore, we find for the Schwarzian
\eq{
	S(f)(z)=\frac{6}{z^2}\sum\limits_{m,n}
	\frac{\pa^2\mathcal{F}}{\pa t_m\pa t_n}z^{-m-n}.
\label{schwarzian-tau}
}
Thus, the expression \eqref{aharon-3} entering the entanglement entropy of the excited states has also representation in terms of dispersionless Toda tau function
\eq{\label{kernel-tau}
	\frac{f'(z)f'(w)}{(f(z)-f(w))^{2}}\,-\,\frac{1}{(z-w)^{2}}=\sum _{m,n}
\frac{\pa^2\mathcal{F}}{\pa t_m\pa t_n}z^{-m-1}w^{-n-1}.
}
Substitution into \eqref{ex-contrib-1} gives
\eq{
S_{vac}-S_{ex}=\frac{c}{12}\log\left(1+(z-w)^2\sum _{m,n}
\frac{\pa^2\mathcal{F}}{\pa t_m\pa t_n}z^{-m-1}w^{-n-1}
\right).
\label{schwarzian-tau-fin}
	}
This expression is one of our main results.

We conjecture  that there is deeper sophisticated relations between integrable hierarchies and entanglement entropies in holography\footnote{See, for instance comments in \cite{deBoer:2016bov}.}. On the other side the latter are main ingredients of information spaces, for instance they are related to Fisher metric etc. 


\section{Conclusions}

In this short note we revisit the problem of contributions of excited states to the entanglement entropy in the context of replica method and one-point correlation function of the energy-momentum tensor in 2d. This problem is of particular importance in holographic dualities, dynamics of excited states as well as in the context of emergent spacetimes. 

The main output of our considerations is following. In Section 2 we represented contributions to the entanglement entropy of excited states through the Aharonov invariants. This result shows that the two-point correlation function of twist operators in replica trick approach gets contributions not only from one-point function of energy-momentum tensor but also from all descendants of the identity operator. The conclusion is that, having given Schwarz-Christoffel map one can easily compute Aharonov invariants to any order and thus, contributions to the entanglement entropy. 
In the next section we use results from the theory of univalent function to relate contributions to the entanglement entropy of excited states to Faber polynomials and represent them in terms of Grunsky coefficients. 
Grunsky coefficients give connection of our problem to the theory of conformal mappings and, concretely to the universal Teichmuller space \cite{sugawa}. One of our main results is the relation of the entanglement entropy of excited states to the tau-function of the dispersionless Toda hierarchy. This relation comes from the representation of Grunsky coefficients as second derivatives of the free energy of dToda hierarchy, or $\log\tau_{dToda}$. It resembles a far analogy of AGT duality \cite{Alday:2009aq} and we expect more insights in this direction.

The results of our study open several obvious directions of research. First of all, one should thoroughly study the physical consequences of these results. In a subsequent paper we will offer some concrete results and particular examples \cite{tobe}. Another issue is the extension of these considerations to the case of other primary operators. The considerations provided in this paper are quite universal and we expect them to be applicable to these cases too. One may also wonder what kind of physics would describes the action in which Schwarzian derivative (see for instance Schwarzian actons in \cite{Maldacena:2016upp}) is replaced by higher Aharonov invariants. The application to the holography, together with quantum Fisher information, emergent geometries etc, is one of the central topics to be investigated within this formalism.

\paragraph{Acknowledgements}\ \\
I thank Daniel Grumiller for critically reading the manuscript. This work was partially supported by the BNSF grant T02/6 and by SU Research Fund grant No 85/2016.


\end{document}